\documentclass{ws-ijmpd}
\usepackage[super,compress]{cite}
\begin{document}

\markboth{Ulrich Sperhake}
{Numerical relativity in higher dimensions}

%
\catchline{}{}{}{}{}
%

\title{NUMERICAL RELATIVITY IN HIGHER DIMENSIONS\footnote{Based on a talk presented at Thirteenth Marcel Grossmann Meeting on General Relativity, Stockholm, July 2012.}
}

\author{ULRICH SPERHAKE
}

\address{Department of Applied Mathematics and Theoretical Physics \\
University of Cambridge, Cambridge CB3 0WA, United Kingdom \\[5pt]
Institute of Space Sciences, CSIC-IEEC, 08193 Bellaterra, Spain \\[5pt]
California Institute of Technology, Pasadena, CA 91125, United States \\[5pt]
CENTRA, Instituto Superior T{\'e}cnico, 1049-001 Lisboa, Portugal\\ [5pt]
U.Sperhake@damtp.cam.ac.uk
}

%

\maketitle

\begin{history}
\received{30 December 2012 }
\revised{--- 2013}
\end{history}

\begin{abstract}
We present an overview of recent developments in numerical relativity
studies of higher dimensional spacetimes with a focus
on time evolutions of black-hole systems.
After a brief review of the numerical
techniques employed for these studies, we summarize results
grouped into the following three areas: (i) Numerical
studies of fundamental properties of black holes,
(ii) Applications of black-hole
collisions to the modeling of Trans-Planckian scattering, (iii)
Numerical studies of asymptotically anti-de Sitter spacetimes in
the context of the gauge-gravity duality.
\end{abstract}


\ccode{PACS numbers: 04.25.D-, 04.70.-s, 04.50.+h}


\section{Introduction}

Numerical modeling of dynamic spacetimes containing black holes (BHs)
in the framework of Einstein's theory has been motivated for a
large part of its nearly 50 year long history by the calculation
of the gravitational wave (GW) signals generated by BH binary
systems. These efforts culminated in 2005 in the first
evolutions of BH binaries
through inspiral, merger and
ringdown\cite{Pretorius:2005gq,Baker:2005vv,Campanelli:2005dd}.
Over the following years, numerical relativity (NR)
simulations have been used in the construction of
GW template banks
\cite{Ajith:2007qp,Ajith:2007kx,Ajith:2011ec,Pan:2011gk,Damour:2010zb}
fundamental for signal searches in GW detectors such as
LIGO, VIRGO and GEO600 \cite{LIGOweb,advancedVIRGO,GEO600web}
and have also been studied
in the context of GW data analysis
\cite{Aylott:2009ya,Aylott:2009tn,Ajith:2012tt}.
The {\em Ninja}\cite{Ninja} and {\em NRAR}\cite{NRAR} collaborations
in particular
combine the resources of various NR groups
with the analytic relativity and gravitational-wave data analysis
community. NR studies
have furthermore provided valuable insight into a variety of
astrophysical scenarios involving BHs as for example
the study of gravitational recoil
\cite{Gonzalez:2006md, Campanelli:2007ew, Gonzalez:2007hi,
Lousto:2011kp},
jet phenomena \cite{Palenzuela:2009hx,Palenzuela:2010nf} and electromagnetic
counterparts for multi-messenger astrophysics
\cite{Farris:2009mt,Neilsen:2010ax,Bode:2011tq,Moesta:2011bn,
Giacomazzo:2012iv,Noble:2012xz}.
For reviews of these subjects we refer to
Refs.~\refcite{Hannam:2009rd,Centrella:2010mx,Hinder:2010vn,Sperhake:2011xk,Pfeiffer:2012pc}.

In recent years, however, NR has also emerged as
a valuable tool to explore high-energy
physics scenarios \cite{Cardoso:2012qm}. The purpose of this article is
to provide an overview of the most recent developments in these areas
which, for reasons of clarity, we divide
into three categories: (i) numerical studies of fundamental properties
of BH spacetimes, (ii) applications to the modeling of Trans-Planckian
scattering and (iii) the gauge gravity duality. This division
is not entirely rigorous and we will encounter scenarios of relevance
for more than one of these major topics.
For example the stability of rotating BHs
in higher dimensions is discussed in Sec.~\ref{sec:fundamental}
on fundamental properties of BHs
but likely also affects the
signature in the conjectured formation of mini BHs in Trans-Planckian
scattering events. Wherever present, we will emphasize such
significance across our somewhat arbitrarily
chosen distinction.

NR studies in the context of high-energy physics
and fundamental properties of classical general relativity
frequently involve BHs in higher dimensional spacetimes.
In some cases, however, valuable insight can be inferred as well
from the more traditional case of spacetimes with ``3+1'' dimensions
(three spatial and one time-like). Somewhat in breach of its title,
this work is therefore not entirely limited to the numerical
investigation of higher-dimensional BHs. As complementary
reading to this article, we note review articles including
NR applications in high-energy physics
by Pretorius\cite{Pretorius2007a}, Cardoso {\em et al.}\cite{Cardoso:2012qm}
and Yoshino \& Shibata\cite{Yoshino:2011zz,Yoshino:2011zza}.
An extended discussion of BHs in higher dimensions is given in
\refcite{Horowitz2012}.

The remainder of this article is organized as follows.
We begin with a brief summary of the NR techniques
in Sec.~\ref{sec:NR}.
The above mentioned main applications
of NR to higher-dimensional BHs
will be reviewed in
Secs.~\ref{sec:fundamental}, \ref{sec:transplanckian}
and \ref{sec:AdSCFT}. In each of these
we will provide a brief motivation of the corresponding
studies and list more specialized review articles for further reading.
We will conclude in Sec.~\ref{sec:conclusions}.
Unless noted otherwise, we use units such that the speed of light $c=1$.

\section{Numerical techniques in higher dimensions}
\label{sec:NR}

Formulations of the Einstein equations suitable for NR
can be classified into two main
groups. The characteristic
Bondi-Sachs\cite{Bondi:1962px,Sachs:1962wk} formalism is based on
the characteristic surfaces of the vacuum Einstein equations.
This formalism leads to a natural
hierarchy of the equations which reflects the isolation of the
gravitational degrees of freedom; in the four-dimensional case,
for example, one obtains one trivial equation, three complementary
equations, four hypersurface equations and two evolution equations
for the degrees of freedom. A difficulty arising in the characteristic
approach is the potential formation of caustics and the ensuing
breakdown of the coordinate system. For the classical general-relativistic
two-body problem, the inspiral and merger of two BHs,
a robust solution for this problem has not
yet been found. Nevertheless, characteristic methods have been
employed with great success in various codes studying configurations
with additional symmetries, most notably in the series of studies
by Chesler \& Yaffe\cite{Chesler:2008hg,Chesler:2009cy,Chesler:2010bi}
on thermalization of quark-gluon plasmas through the
Anti-de Sitter/Conformal Field Theory (AdS/CFT) correspondence.
Characteristic evolutions are reviewed in detail by
Winicour\cite{Winicour2012}.

The majority of work on the inspiral and coalescence of compact binaries,
including the 2005
breakthroughs,
however, has been performed in 3+1 approaches. Here spacetime
is decomposed into a one-parameter family of three or,
more generally, $d-1$ dimensional, spatial hypersurfaces. The
canonical 3+1 split has been developed by
Arnowitt, Deser and Misner\cite{Arnowitt:1962hi} (ADM),
reformulated by York\cite{York1979} and leads to a system of
six second-order evolution equations for the induced spatial
metric and four constraint equations, the Hamiltonian and
momentum constraints. Introduction of the extrinsic curvature
as an evolution variable converts the evolution equations into
a first-order-in-time system. By performing a conformal rescaling of
the spatial metric, a decomposition of the extrinsic curvature
into trace and a trace-free part and evolving the contracted
Christoffel symbol as separate fields,
Baumgarte, Shapiro, Shibata and Nakamura\cite{Baumgarte:1998te,Shibata:1995we}
have developed a strongly hyperbolic formulation\cite{Gundlach:2006tw}
commonly referred to as ``BSSN'' which has been used with great success
by many numerical groups. The ADM/BSSN approach to numerical relativity
is discussed in great
detail in \refcite{Gourgoulhon:2007ue,Alcubierre2008,Baumgarte2010}.

An alternative Cauchy-type formulation
has been derived from the Einstein equations in harmonic
gauge where they take on a manifestly hyperbolic form.
The harmonic gauge can be generalized to arbitrary gauge
choices by introducing source functions for the coordinate conditions
\cite{Friedrich1985,Garfinkle:2001ni}. More details of this {\em generalized
harmonic gauge} (GHG) formulation can be found in
\refcite{Pretorius:2004jg,Lindblom2005,Gundlach:2005eh}.
The GHG and BSSN methods have been the foundation of the numerical
relativity breakthroughs of 2005 by
Pretorius\cite{Pretorius:2005gq} and the Brownsville and Goddard
groups\cite{Campanelli:2005dd,Baker:2005vv}.

NR applications in higher-dimensional spacetimes
have so far been restricted to configurations with
symmetry assumptions. This is largely a consequence
of the vast increase in computational resources required by
each additional dimension. A reduction of higher-dimensional
general relativity to an effective 3+1 or lower-dimensional
computational domain has been achieved in various ways. The most
evident approach is to directly impose the symmetry together with
the coordinate conditions on the line element.
For example, Sorkin \& Oren\cite{Sorkin:2005vz} describe
the $d$-dimensional, spherically symmetric
spacetime in double-null coordinates $(u,v)$
by the line element
\begin{equation}
  ds^2 = -\alpha(u,v)^2 du\,dv + r(u,v)^2 d\Omega_{d-2}^2\,,
\end{equation}
where $d\Omega_{d-2}$ is the metric on a $d-2$ dimensional unit hyper sphere
and $r$ is the areal radius. The time evolution is then determined
by the Einstein equations derived from this metric; see their Eqs.~(3)-(6).

Alternatively, a general procedure for a dimensional reduction
under the assumption of one or more Killing vectors is
given by Geroch\cite{Geroch:1970nt}. By using this procedure,
the $d$-dimensional Einstein equations with $SO(n)$ symmetries
can be converted to an effective 3+1 scheme coupled to
scalar or vector fields which represent the curvature of the
extra-dimensions\cite{Sorkin:2009wh,Zilhao:2010sr}.
Such a dimensional reduction is particularly attractive because
it provides a straightforward way to apply the
successful numerical techniques developed in 3+1 with
relatively minor modifications to the study of
higher dimensional spacetimes.

A different way to achieve the same goal is based on the
{\em cartoon method}\cite{Alcubierre:1999ab} which employs
interpolation of grid functions and/or trading of spatial
derivatives to evolve symmetric spacetimes
on a reduced Cartesian grid. This approach has been modified and
extended to higher-dimensional applications by
Pretorius\cite{Pretorius:2004jg} and
Yoshino \& Shibata\cite{Yoshino:2009xp,Yoshino:2011zza}
and has proven
a rather robust tool to study BH collisions and instabilities.

The generation of initial data in higher-dimensional spacetimes
relies to a significant extent on similar techniques developed
for the 3+1 case; see Cook's\cite{Cook2000} review.
In particular, it has been shown by
Yoshino {\em et al.}\cite{Yoshino:2006kc} that the existence of analytic
solutions of Bowen-York\cite{Bowen:1980yu} type for the momentum
constraints carries over to general relativity in $d$ dimensions.
Solving the Hamiltonian constraint for the conformal factor
can then be achieved by numerical methods similar to those
developed for 3+1 dimensions. By using this procedure, initial
data for BH binaries with boost have been obtained in
Refs.~\refcite{Yoshino:2006kc,Zilhao:2011yc}. For specific numerical
studies, analytic initial data is available in closed form, as for
example the $d$ dimensional generalization of Brill-Lindquist\cite{Brill1963}
data describing a BH binary at the moment of time symmetry.
Analytic BH solutions can also be employed to construct approximate
solutions to the Einstein constraints by superposition of
either the solutions themselves or boosted variants thereof.

For more details on the numerical methods, the construction of initial
data and gauge choices, we refer the reader to the publications
reviewed in the following sections.

\section{Fundamental properties of black holes}
\label{sec:fundamental}

BHs play a central role in the applications of NR
to higher-dimensional spacetimes discussed in
Secs.~\ref{sec:transplanckian} and \ref{sec:AdSCFT}.
Quite aside from these applications,
BHs provide a unique probe for deepening our insight
into the gravitational theory and understanding in which regards
3+1 dimensions represent a special case. BHs in
higher dimensions indeed reveal a much richer phenomenology
than their four-dimensional cousins;
cf.~Emparan \& Reall\cite{Emparan:2006mm,Emparan:2008eg}.

Asymptotically flat, stationary solutions of the Einstein
equations in vacuum in 3+1
dimensions belong to the Kerr family and describe single
BHs of spherical topology
characterized by two parameters, the mass $M$ and
angular momentum parameter $a$
\cite{Hawking1980,Heusler:1998ua,Chrusciel:2012jk}.
The stability of these solutions
has been studied extensively;
see for example Refs.~\refcite{Kay:1987ax,Whiting:1988vc} and
Dafermos'\cite{Dafermos:2008en} review. Such classical BH solutions
are expected to be stable in the region outside the horizon,
but admit instabilities in the interior\cite{Dotti:2011eq}.
Penrose's {\em cosmic censorship}\cite{Penrose1978,Penrose:1999vj}
furthermore conjectures that
collapse of ordinary matter does not lead to the formation of naked
singularities; instead such singularities are hidden from outside
view through an event horizon.

BHs in higher dimensions are solutions to the $d$-dimensional
Einstein equations\footnote{see Sec.~3.1 in \refcite{Emparan:2008eg} for
a discussion of Newton's constant $G$}
\begin{equation}
  G_{\mu \nu} = R_{\mu \nu} - \frac{1}{2} g_{\mu \nu} R = 8\pi G T_{\mu \nu},
  \label{eq:Einstein}
\end{equation}
for the vacuum case $T_{\mu \nu}=0$.

Solutions to these equations can
be generated straightforwardly from extending the four-dimensional
Schwarzschild metric with one or more flat dimensions. These
{\em black strings} or {\em black branes} are susceptible to the
Gregory-Laflamme\cite{Gregory:1993vy,Gregory:1994bj} (GL)
instability related to the vastly different length scales in the
horizon geometry. These solutions furthermore suggest a modification of
Thorne's\cite{Thorne:1972ji} Hoop conjecture; if an amount $M$ of
mass-energy is compressed into a volume such that its circumference
is below its Schwarzschild radius in every direction, a BH forms.
The existence of black branes indicates that such a concentration of
energy is sufficient in a subset of the spatial directions.

The $d$-dimensional generalization of
the spherically symmetric Schwarzschild solution is given by the
Tangherlini\cite{Tangherlini:1963bw} metric
\begin{equation}
  ds^2 = -\left( 1-\frac{\mu}{r^{d-3}} \right) dt^2
         + \left(1-\frac{\mu}{r^{d-3}} \right)^{-1} dr^2
         + r^2 d\Omega_{d-2}^2,
\end{equation}
where the constant $\mu$ determines the BH mass
$M=(d-2) \Omega_{d-2} \mu / (16\pi G)$. Here, $\Omega_{d-2}$ denotes the
surface area of the $d-2$ dimensional unit hypersphere.
Note that mass is now of dimension
${\rm length}^{d-3}$. Schwarzschild solutions in $d$ dimensions have been
found to be stable against linearized gravitational
perturbations\cite{Gibbons:2002pq,Ishibashi:2003ap}.

In contrast to their spherically symmetric counterparts, the class
of rotating BH solutions exhibits a vastly richer structure
in higher dimensions. First, rotation is now possible in more than
one plane, giving rise to more than one rotation parameter. Second,
the radial dependence of the Newtonian potential is altered to
$1/r^{d-3}$ whereas the centrifugal potential maintains its
$1/r^2$ character.
This $d$-dependent character of the Newtonian potential
manifests itself in the absence of stable circular geodesics for
$d>4$\cite{Cardoso:2005vk,Cardoso:2008bp}.

Quite remarkably, Myers \& Perry\cite{Myers:1986un}
found an exact solution describing BHs rotating in all possible
planes.
In $d=4$ and $d=5$, BHs rotating in one plane with
rotation parameter $a>M$ and $a^2 > \mu$ respectively represent naked
singularities. In $d\ge 6$, in contrast, there exists a horizon for all
values of $a$. Furthermore, in the limit of infinite rotation rate,
this horizon becomes increasingly flattened
along the rotation plane\cite{Emparan:2003sy} and is therefore
expected to be GL unstable. This has been confirmed in linearized
studies by Dias {\em et al.}\cite{Dias:2009iu,Dias:2010eu}.
In the case of rotations in more than one plane, the
existence of horizons in ultra-spinning BH spacetimes requires
at least two (one) spin parameter to vanish for an odd (even) number
of dimensions $d$.

Rotation in higher dimensions also leads to BH solutions
of non-spherical topology. {\em Black rings} are supported against
contraction by the centrifugal potential; analytic solutions
in $d=5$ have been found in \refcite{Emparan:2001wn,Pomeransky:2006bd}
Thin rings behave locally like
boosted black strings\cite{Elvang:2003mj} which are
GL unstable\cite{Hovdebo:2006jy} while {\em fat} rings
are expected to be unstable to variations of their
radius\cite{Elvang:2006dd}. Solutions for {\em black saturns}
have been constructed in \refcite{Elvang:2007rd} and demonstrate the
existence of stationary vacuum multiple-BHs in higher dimensions.

The generalization of the AdS-Schwarzschild solution\cite{Kottler1918}
is given by
\begin{equation}
  ds^2 = -\left( 1-\frac{\mu}{r^{d-3}} + \frac{r^2}{L^2} \right) dt^2
         + \left( 1-\frac{\mu}{r^{d-3}} + \frac{r^2}{L^2} \right)^{-1} dr^2
         + r^2 d\Omega_{d-2}^2.
\end{equation}
In the limit $\mu \rightarrow 0$, this solution reduces to
the AdS spacetime with curvature radius $L$ in {\em global} coordinates;
cf.~Sec.~\ref{sec:AdSCFT}. Its mass relative to the empty AdS solution is
given by \cite{Ashtekar:1984zz,Ashtekar:1999jx}
$M=(d-2)\Omega_{d-2} \mu \ (16\pi G)$. This solution is
stable against linearized gravitational perturbations
\cite{Kodama:2003jz}. The four-dimensional Kerr-AdS spacetime
(see e.~g.~Eq.~(2.1) in \refcite{Dias:2012pp})
has been found by Carter\cite{Carter:1968ks}
and generalized to higher dimensions in
Refs.~\refcite{Hawking:1998kw,Gibbons:2004uw,Gibbons:2004js}.
Kerr-AdS has been found to be unstable against linear perturbations
if the BHs are small\cite{Hawking:1999dp,Cardoso:2004hs,Cardoso:2006wa,
Kodama:2009rq}.

One of the first and most influential NR results has been obtained in
Choptuik's\cite{Choptuik:1992jv} seminal study of the collapse
of spherically symmetric massless scalar fields minimally coupled
to Einstein gravity in four-dimensional, asymptotically flat spacetimes.
By evolving various one-parameter families of scalar pulses,
he identified critical behaviour as the parameter
$p$ which characterizes the gravitational interaction strength
of the field approaches a critical value $p^*$.
For $p>p^*$, the field collapses to a BH and for $p<p^*$
it disperses to infinity. Furthermore, near-critical field configurations
exhibit {\em universal} behaviour in the strong field limit:
(i) BHs which form have a mass $M\sim |p-p^*|^{\gamma}$ with a
universal constant $\gamma \approx 0.37$. (ii) Advancing the evolution
from a time $t$ to $t + \Delta$, the field profile is recovered
up to a ``zoom-in'' by a factor $e^{\Delta}$. The numerical study
reveals $\Delta$ to be a universal constant of about $3.4$.
For subcritical configurations,
Garfinkle \& Comer Duncan \cite{Garfinkle:1998va} found
a similar scaling of the maximal scalar curvature in the spacetime
$R_{\rm max}\sim |p-p^*|^{2\gamma}$. Continuously self-similar
solutions were found by Pretorius \& Choptuik\cite{Pretorius:2000yu}
for scalar fields in 2+1 dimensional asymptotically AdS spacetimes
with a mass scaling characterized by $\gamma/2 = 1.2$.
Sorkin \& Oren\cite{Sorkin:2005vz} generalized Choptuik's result
to higher dimensions by evolving scalar fields
up to $d=11$ dimensions. Their results indicate that $\gamma$
reaches a maximum and $\Delta$ a minimum around $d \approx 11\ldots 13$.
An extended review of critical collapse phenomena is given by
Gundlach \& Mart{\'i}n-Garc{\'i}a\cite{Gundlach:2007gc}.

More recently, critical-collapse studies have been extended to
asymptotically AdS spacetimes in 3+1 and higher dimensions.
In a remarkable study, Bizo{\'n} \& Rostworowski found evidence
suggesting that the 3+1 AdS is unstable to BH formation
under arbitrarily small perturbations. By evolving spherically
symmetric scalar fields
they recover Choptuik's results for large initial field amplitudes.
For configurations below this critical value, however, the
AdS boundary substantially modifies the outcome. In contrast to
asymptotically flat spacetimes, spatial infinity in AdS is reached in
finite time by massless fields and reflected back onto the
origin. As the initial amplitude is reduced below
a critical value $\epsilon_0$, the scalar pulse forms a BH upon its second
implosion on the origin. Further reduction of the amplitude leads to
a second critical amplitude $\epsilon_1$ and this pattern repeats
itself with no indication of a threshold amplitude for
BH formation; cf.~their Fig.~1.
For each critical amplitude, they
furthermore recover Choptuik's scaling law with $\gamma=0.37$.
They interpret this behaviour as a resonant mixing of modes which
transfers energy from low to high frequencies.
This study has been generalized to higher-dimensional AdS spacetimes
by Ja{\l}mu\.{z}na {\em et al.}\cite{Jalmuzna:2011qw} suggesting
that AdS is unstable to BH formation for generic spacetime dimensions.
Presumably, this result has eluded a
similar study by Garfinkle \& Pando Zayas\cite{Garfinkle:2011hm},
because of insufficient length of their numerical simulations
for smaller field amplitudes.
In a similar investigation using complex scalar fields,
Buchel {\em et al.}\cite{Buchel:2012uh} reproduce the instability
of AdS and the transfer of energy from low to high frequencies. In
consequence, the width of initially weak pulses narrows in each
reflection cycle and eventually collapses to a BH; cf.~their Fig.~5.
Garfinkle {\em et al.} \cite{Garfinkle:2011tc} have monitored the
time of BH formation and the horizon radius and find the
amplitude of the scalar field to have a stronger influence on the
outcome compared with the width of the pulse. The same type of instability
to BH formation has been found by Maliborski \cite{Maliborski:2012gx}
for a Minkowski spacetime enclosed inside a reflecting wall, indicating that
the global structure plays a major role for the effect.
Perturbative studies support the numerically observed instability
of AdS \cite{Dias:2011ss}, but the question of the generic stability
properties of {\em asymptotically} AdS spacetimes,
as for example those containing BHs or boson stars, remains
under study \cite{Dias:2012tq}.

The instability of black strings
has been studied numerically in
a sequence of papers by Lehner \& Pretorius and
collaborators\cite{Choptuik:2003qd,Garfinkle:2004em,Lehner:2010pn},
see also \refcite{Lehner:2011wc}.
It had been known since the 1990s that black strings are
subject to the GL instability, but the eventual fate of the string remained
unclear. In Ref.~\refcite{Lehner:2010pn}, Lehner \& Pretorius
found evidence supporting indications by earlier work that the
string evolves to a sequence of spherical BHs connected by thin
string segments which themselves are subject to the GL instability,
resulting in a self-similar cascade reaching zero string width
in finite asymptotic time. This behaviour shows striking similarity
with satellite formation in the flow of low-viscosity fluids.
The eventual bifurcation of the horizon resulting from the
cascade implies formation of a naked singularity\cite{Hawking1973}.
Because no finetuning is required to trigger the instability,
the result constitutes a violation of the cosmic censorship
conjecture without ``unnatural'' assumptions about the initial data.
In contrast to the higher-dimensional case, NR has as yet not observed any
such violation of cosmic censorship in 3+1 dimensions. In a
recent study\cite{Zilhao:2012bb}
BH collisions in asymptotically de Sitter spacetimes have been found
to comply with censorship; BH binaries with a combined mass exceeding
the inverse of the Hubble constant do not merge for any initial
separation provided the initial data do not contain a naked singularity.

We have already mentioned the linear instability of rapidly rotating
Myers-Perry BHs. The stability of rotating BHs has been studied numerically
by Yoshino \& Shibata\cite{Shibata:2009ad,Shibata:2010wz}
who evolve single holes with a bar-shaped perturbation
imposed on the conformal factor in $5\le d \le 8$ dimensions.
In all cases, they observe BHs spinning above a threshold
dimensionless spin-parameter $j_{\rm crit}$
to spontaneously emit gravitational waves and settle down into
a subcritical configuration. Expressed in terms of the ratio of
polar to equatorial circumference $C_p/C_e$ of the apparent horizon,
the onset of instability occurs for nearly identical
values for $d\ge 6$, namely $0.65$, $0.68$ and $0.67$, respectively, for $d=6$,
$7$ and $8$. In contrast, they obtain a significantly
stronger distortion threshold $0.38$ in $d=5$ indicating a somewhat
special status of $d=5$; we recall in this context the absence of
an upper limit on the spin in $d\ge 6$. BHs spinning below the
critical rate, on the other hand, show no signs of instability
in any of their simulations. The effect of this instability
on the evaporation of BHs conjectured to be produced in
trans-Planckian scattering (cf.~Sec.~\ref{sec:transplanckian})
depends on the timescale of the instability and, hence, on the
spin parameter of the formed BH; for details see the
discussion in Sec.~VI D in Ref.~\refcite{Shibata:2010wz}.

Even though this review is primarily concerned with time
evolutions, we note that numerical techniques are also applied
for finding static and stationary solutions to the
$d$ dimensional Einstein equations in vacuum, as for example
the construction of four-dimensional rotating BH solutions in a
cavity\cite{Adam:2011dn} or asymptotically AdS spacetimes in $d=5$
with a conformally Schwarzschild boundary \cite{Figueras:2011va}.
This topic is reviewed in detail by Wiseman\cite{Wiseman:2011by}.

\section{Trans-Planckian scattering}
\label{sec:transplanckian}

The standard model of particle physics provides an exceptionally successful
description of subatomic particles and their interactions via
the electromagnetic, weak and strong forces. In spite of its success,
however, there remain important unanswered questions, as for example
the unknown nature of dark matter. In the context of this section,
the most important open issue is the
large discrepancy between the electroweak
energy scale $246~{\rm GeV}$ and the grand
unification scale $\sim 10^{19}~{\rm GeV}$.
This so-called {\em hierarchy problem}
manifests itself in the extraordinary weakness of the gravitational
interaction relative to the other fundamental forces; the weak
interaction is about 32 orders of magnitude stronger than
the gravitational one. At energies comparable to the grand unification
energy, on the other hand, gravity is expected to become comparable
in strength to the other interactions.

A particularly intriguing solution to the
hierarchy problem is the {\em ADD} model
\cite{Antoniadis:1990ew, ArkaniHamed:1998rs, Antoniadis:1998ig},
which proposes large
(relative to the four-dimensional Planck scale $M_{\rm Pl}$)
extra dimensions of radius $\sim R$
such that the fundamental Planck scale
$M_{{\rm Pl},d}$ may be as low as $\sim 1~\mathrm{TeV}$. Consider
for illustration the Poisson equation in $d-1$ spatial dimensions
for a point source $\Delta \Phi=4\pi G_dM$, with the $d$ dimensional
gravitational constant $G_d=\hbar^{d-3}c^{5-d}M_{{\rm Pl},d}^{2-d}$
and the source mass $M$.
From Gauss's law we obtain in spherical coordinates
\begin{equation}
  I_d := \int \Delta \Phi dV = \int \nabla \Phi d \mathbf{A}
      = \partial_r \Phi r^{d-2}\Omega_{d-2}
      = \partial_r \Phi \frac{2\pi^{(d-1)/2}}{\Gamma(\frac{d-1}{2})}r^{d-2}
      = 4\pi G_d M\,.
\end{equation}
At large distances $r\gg R$, the extra dimensions are not
accessible to the gravitational field and merely contribute an overall
volume factor such that $I_d \sim R^{d-4} I_4$ whereas for $r<R$, the
potential is determined by the $d-1$ dimensional Poisson equation. Ignoring
geometrical factors of order $\mathcal{O}(1)$, we thus obtain
\begin{equation}
  \Phi \sim -\frac{G_d M}{r^{d-3}}~~\mathrm{for}~~r<R\,,~~~~~~~~~~
  \Phi \sim -\frac{G_d M}{R^{d-4} r}~~\mathrm{for}~~r\gg R\,.
\end{equation}
At large distances we thus recover the usual $1/r$ potential with
a four-dimensional coupling constant $G_4 = G_d / R^{d-4}$
which implies for the energy scales $M_{\rm Pl}^2 \sim
M_{{\rm Pl},d}^{d-2} R^{d-4}$. For example, a fundamental Planck mass
$M_{{\rm Pl},d}$ of the order of the electroweak scale requires
$R\sim 10^{13}~{\rm cm}$ in $d=5$ (ruled out experimentally) or
$R \lesssim 1~{\rm mm}$ for $d=6$.
In a different version of the model proposed by
Randall \& Sundrum\cite{Randall:1999ee,Randall:1999vf},
a finite length scale is introduced through a warp factor
into otherwise infinite extra dimensions.

An intriguing consequence of an effective Planck mass much below
its four-dimensional value $\sim 10^{19}~{\rm GeV}$ is the possibility
of BH formation in parton-parton collisions at the
Large Hadron Collider (LHC) \cite{Dimopoulos:2001hw,Giddings:2001bu};
see also the reviews by Cavagli{\`a} \cite{Cavaglia:2002si}
and Kanti \cite{Kanti:2008eq}.
Once formed, such mini BHs are expected to evaporate in
four stages \cite{Kanti:2008eq}: (i) a balding phase during which the
BH sheds all multipoles except for mass, spin and charge, (ii) a
spin-down and (iii) a Schwarzschild phase during which the BH looses first
its spin and then its mass via semi-classical
Hawking radiation and (iv) the Planck
regime as the BH mass approaches the Planck mass which is
described by an as yet unknown theory of quantum gravity.

Of particular interest in the context of NR is the fact that the first
three of these phases should be well described by classical and
semi-classical calculations provided the BH mass exceeds the Planck
scale by at least a factor of a few \cite{Kanti:2008eq}. If we
further assume that most of the energy of the collision process resides
in the kinetic energy of the particles, such that their internal
structure becomes negligible, the dynamics of the collision should be
well modeled by two point particles or BHs in $d$-dimensional
general relativity \cite{Banks:1999gd, Giddings:2001bu}. Testing this
assumption forms one of the main motivations for NR applications and
we shall return to this question shortly.

BH formation is expected to manifest itself in collision experiments
by a special signature in its decay products as for example the
jet multiplicity or transverse energy \cite{Chatrchyan:2012taa}.
For the identification of these signatures, theoretical predictions from
Monte-Carlo event generators such as {\sc BlackMax}
\cite{Dai:2007ki} and {\sc Charybdis}
\cite{Harris:2003db,Frost:2009cf} are compared with experimental
data. Key input parameters for the event generators are the
scattering cross section for BH formation and the initial mass and
spin distributions of the formed holes. Providing this information
forms a second main challenge for NR to complement estimates
obtained from perturbative calculations on superposed
Aichelburg-Sexl shock-wave backgrounds
\cite{D'Eath:1976ri,D'Eath:1992hb,D'Eath:1992hd,Yoshino:2005hi,
Herdeiro:2011ck,Coelho:2012sya,Coelho:2012sy}.

The most recent analysis of data taken from collisions at the LHC with
energies up to $8~{\rm TeV}$ excludes semi-classical BHs of mass below
$4.1-6.1~{\rm TeV}$ \cite{CMS2012} and sets exclusion contours in the
plane formed by the BH mass and the threshold mass separating the
semi-classical from the quantum regime \cite{ATLAS2012}. It thus appears
unlikely that the most optimistic TeV-gravity models with
$M_{{\rm Pl},d}\approx 1~{\rm TeV}$ are those realized in nature.

In summary, the main questions to be addressed by NR simulations
are (i) the validity of the Hoop conjecture in highly dynamical configurations,
(ii) the scattering threshold for BH formation, (iii) the mass and spin
parameters of the formed BHs and
(iv) the impact of the colliding objects' structure on the
dynamics.

High-energy collisions in the framework of general
relativity are currently best understood in $d=4$ dimensions, largely
because the numerical infrastructure for this setting is most advanced and
robust. Even though generalization of the results to arbitrary spacetime
dimensionality will ultimately be essential, a great deal of insight
can be and has already been obtained in the four-dimensional case. We
shall review these results before plunging into the latest developments
on collisions for arbitrary $d$.

The expectation of BH formation in high-energy collisions dominated by
gravitational interaction is closely tied to Thorne's Hoop conjecture
\cite{Thorne:1972ji}.
Applied to head-on collisions with boost factor $\gamma$,
we expect two objects, each of rest mass $m_0$ and size $R_0$,
to form a BH if the Schwarzschild radius $R=2~M=4\gamma m_0$
associated with $M$ satisfies $R> R_0$, i.~e.~$\gamma > R_0/(4m_0)$.
Choptuik \& Pretorius
\cite{Choptuik:2009ww} tested this scenario by numerically simulating
head-on collisions of boosted boson stars in $d=4$ dimensions. Their
initial configurations consist of two boson stars with compactness
$2m_0/R_0 \approx 1/20$ and the Hoop conjecture would imply BH formation
for boost factors $\gamma > 10$. Their numerical results demonstrate
BH formation for $\gamma > 2.9\pm 10~\%$, a factor a few below the
Hoop conjecture and thus strengthening the expectation that kinetic-energy
dominated collisions indeed produce BHs. This picture has recently been
confirmed in head-on collisions of fluid particles in
Refs.~\refcite{East:2012mb,Rezzolla:2012nr}. The energy
radiated in gravitational waves (GWs)
reported by East \& Pretorius\cite{East:2012mb} is
$\sim 16\pm2~\%$ for $\gamma=10$, in good agreement
with ultrarelativistic limits
$16.4~\%$ and $14\pm3~\%$ obtained from perturbative
calculations \cite{D'Eath:1976ri,D'Eath:1992qu}
and numerical simulations of BH head-on collisions \cite{Sperhake:2008ga}
respectively. These studies strengthen the conjecture that trans-Planckian
collisions are well described by high-energy collisions of black holes,
general relativity's closest analog of point particles.

Binary configurations of non-spinning, equal-mass BH binaries are
characterized by two parameters, the boost factor $\gamma=1/\sqrt{1-v^2}$
and the impact parameter defined as $b=L/P$, where $v$, $L$ and $P$
are the velocity, orbital angular momentum and linear momentum.
The head-on case $b=0$ has been studied by Sperhake {\em et al.}
\cite{Sperhake:2008ga} who evolve a sequence of collisions with
$\gamma \le 2.93$. The simulations always result in a merger and
extrapolation of the radiated energy to $v\rightarrow 1$ leads to the
above mentioned $E_{\rm rad}=14\pm 3~\%$ of the Center-of-Mass
(CoM) energy, about a
factor 2 below Penrose's upper limit $1-1/\sqrt{2}$ quoted in
\refcite{Eardley:2002re} . For grazing collisions, the outcome depends
on the size of the impact parameter $b$.
Sperhake {\em et al.}\cite{Sperhake:2009jz}
identify three regimes: (i) For $b\ge b_{\rm scat}$, the BHs scatter
off each other and eventually escape to infinity. (ii) In a narrow
regime $b* \le b < b_{\rm scat}$ the BHs separate after an initial
encounter, but have lost enough energy in gravitational radiation
such that they form a bound system and eventually merge. As
$b$ approaches the {\em threshold of immediate merger}
$b^*$, the binary exhibits the {\em zoom-whirl} behaviour
identified by Pretorius \& Khurana\cite{Pretorius:2007jn} with a number
of whirls proportional to the logarithmic distance of $b$ from $b^*$.
(iii) For sufficiently small $b<b^*$, the outcome is a prompt merger.
By fitting numerical results
obtained for binaries with $\gamma \le 2.3$, Shibata {\em et al.}
\cite{Shibata:2008rq} have obtained a functional relation for the scattering
threshold
\begin{equation}
  b_{\rm scat} = \frac{2.5 \pm 0.05}{v} M.
\end{equation}
The numerical studies further
demonstrate that grazing collisions with impact parameters
near the range $b^* \lesssim b_{\rm scat}$ generate enormous amounts of
gravitational radiation up to at least $\sim 35~\%$ of the CoM energy;
for a detailed analysis of the GW emission see
Berti {\em et al.}\cite{Berti2010}.
Indeed it has been conjectured by Pretorius \& Khurana
\cite{Pretorius:2007jn} that {\em all}
kinetic energy may be radiated in the form of GWs. This conjecture
has recently been investigated numerically in
Sperhake {\em et al.} \cite{Sperhake:2012me} by monitoring the
BH horizon properties in collisions of spinning and non-spinning
binaries. While most of the kinetic energy is indeed radiated
for mild boosts $\gamma \lesssim 1.5$, GW absorption by the
holes becomes increasingly
efficient for larger collision velocities; cf.~their Fig.~1.
Extrapolation of the numerical results to the ultrarelativistic
limit predicts that approximately equal fractions of the kinetic energy
are absorbed by the BHs as are lost in gravitational radiation,
and sets a lower bound on the
final BH mass of about half of the CoM energy. This study
further demonstrates that the impact of the BH spin on the scattering
threshold and radiated energy
becomes negligible for $\gamma \gtrsim 2.5$ lending additional
support to the modeling of trans-Planckian scattering in terms
of high-energy collisions of BHs.

The most likely scenarios for BH formation at the LHC are
proton-proton collisions. It will be important, therefore, to
investigate the impact of electric charge on the dynamics which has
been ignored in the above mentioned studies.
A first exploration of colliding BHs with electric charge has been
presented in Zilh{\~a}o {\em et al.}
\cite{Zilhao:2012gp}. They extract gravitational
as well as electromagnetic radiation generated in head-on collisions
of BHs with equal mass and charge starting from rest. As intuitively
expected, the electric repulsion
slows down the dynamics and leads to a decreasing amount of
GW energy $E_{\rm rad}^{\rm gr}$
as the mass-to-charge ratio $Q/M$ increases from $0$
towards the extremal case $Q=M$ which corresponds to a static,
albeit unstable, configuration. The radiated electromagnetic energy
$E_{\rm rad}^{\rm el}$,
on the other hand, is maximal near $Q/M =0.6$ but always remains
below the gravitational one with the quotient $E_{\rm rad}^{\rm gr}
/E_{\rm rad}^{\rm el}$ increasing monotonically as a function of $Q/M$
reaching $~0.25$ as $Q/M \rightarrow 1$.

In contrast to the four-dimensional case, BH collisions in $d\ge 5$
spacetime dimensions appear to be more challenging from the view point
of numerical stability and have not yet been explored to the same extent.
Collisions of two equal-mass, non-spinning BHs starting
from rest have been studied by Witek {\em et al.}
\cite{Witek:2010xi,Witek2010c} who use the Kodama-Ishibashi formalism
\cite{Kodama:2003jz} for the extraction of GWs
and find that $0.089~\%$ of the CoM energy is radiated in GWs for $d=5$,
to be compared with $0.05~\%$ in four dimensions
\cite{Sperhake:2006cy}.
These results were generalized to the
case of unequal mass ratios $\eta:=m_1 m_2/(m_1 + m_2)^2 <0.25$
in \refcite{Witek2010c} (cf.~their Fig.~3)
leading to the extreme-mass-ratio limit $E^{\rm gr}_{\rm rad}/(M\eta^2)=0.0164
-0.0336\eta^2$.

Point-particle calculations are in good qualitative and
quantitative agreement with these NR results\cite{Berti2010} but additionally
make very interesting
predictions for BH head-on collisions with {\em non-zero} initial
velocities \cite{Berti:2010gx}: $E^{\rm gr}_{\rm rad}$ typically
{\em decreases} with $d$ to such an extent that for $d=11$,
head-on collisions starting from rest may radiate more GW energy than
their ultrarelativistic counterparts; cf.~Fig.~1 in \refcite{Berti:2010gx}.
This counter-intuitive prediction has yet to be tested with NR methods.
BH collisions in $d=5$ dimensional cylindrical spacetimes have been
simulated in \refcite{Zilhao:2011zz}. The cylindrical setup can be
viewed in terms of an infinite array of BHs resulting in a
breaking effect on the collision dynamics.

The most advanced study of BH scattering in $d=5$ has been performed
by Okawa {\em et al.} \cite{Okawa:2011fv} who find scattering
thresholds $b_{\rm scat}$ decreasing
in terms of the Schwarzschild radius $r_{\rm S} = \mu^{1/(d-3)}$
from $~3.6~r_{\rm S}$ at $v=0.4$
to $~3.3~r_{\rm S}$ at $v=0.6$. Above this velocity, numerical
instabilities prevent a determination of $b_{\rm scat}$.
In contrast to the four-dimensional case, however, they
do not observe any signs of zoom-whirl behaviour, probably
due the equal fall-off $\sim1/r^2$ of the gravitational and
centrifugal potential in $d=5$. By analysing the Kretschmann
scalar $C:=R^{\alpha \beta \gamma \delta} R_{\alpha \beta \gamma \delta}$,
they further conclude that super-Planckian physics may be visible
in the scattering of BHs in five spacetime dimensions. A straightforward
calculation shows that for a single BH of mass $M$, the Kretschmann
scalar on the horizon is given by\footnote{
Note the difference of a factor $3\pi/8$ arising from the different
choices of $G$ in the Einstein equations,
our Eq.~(\ref{eq:Einstein}) and Eq.~(1)
in \refcite{Okawa:2011fv}.}
\begin{equation}
  \sqrt{C_{\rm hor}} = \frac{3\pi}{8} 6\sqrt{2}
      \frac{M_{{\rm Pl,}D}^3}{\hbar M}.
\end{equation}
By calculating $\sqrt{C}$ in the orbital plane, Okawa {\em et al.}
observe super-Planckian domains where
$C\gg C_{\rm hor}$ outside the individual holes horizons
even for non-merging configurations. Regions with curvature radii
below the Planck scale may thus form in high-energy particle
collisions without being cloaked inside an event horizon.

\section{Gauge-gravity duality}
\label{sec:AdSCFT}

The Gauge-gravity duality conjectures the mathematical equivalence
or {\em duality}
between field theories involving gravity in $d$ dimensions on the
one side and field theories without gravity in $d-1$ dimensions on the
other. It is one of the most remarkable results of string theory
and, because of Maldacena's
prototypical example
\cite{Maldacena:1997re,Gubser:1998bc,Witten:1998qj}
of the duality between
type IIb string theory in 5 dimensional AdS
space times the $S^5$ sphere and the $\mathcal{N}=4$ Supersymmetric
$SU(N)$ Yang-Mills (SYM) CFT
in 4 dimensions, often referred to as the AdS/CFT correspondence.

Following Maldacena's original example, a large number of similar dualities
have been conjectured; see for example chapter 13 in
Ref.~\refcite{Nastase:2007kj}. Many of these involve type IIb string theory
or $M$ theory compactified on ${\rm AdS}_n$ times a sphere or torus. The
best understood example to date, however, is given by Maldacena's prototype
which is also the version underlying present NR studies and, hence,
the one we shall focus on in the remainder of this section.

The AdS/CFT correspondence, while not rigorously proven, is inspired and
supported by a variety of features displayed by the two theories involved:
\begin{itemize}
  \item Holography: Black holes can be regarded as thermodynamic
        systems with entropy $S=A_{\rm hor}/4G$, where $A_{\rm hor}$
        is the horizon area \cite{Bekenstein:1974ax,Hawking:1976de}.
        In contrast, the entropy of
        systems described by local field theories
        grows in proportion to the volume, suggesting
        a holographic connection to gravity
        \cite{'tHooft:1993gx,Susskind:1994vu,Bousso:2002ju}.
  \item Symmetries: The isometry group of $AdS_5 \times S^5$
        is $SO(2,4) \times SO(6)$ which also leave $\mathcal{N}=4$
        SYM invariant. Fermionic symmetries agree similarly and
        lead to a supergroup $SU(2,2|4)$ which is identical to
        the $\mathcal{N}=4$ superconformal symmetry;
        cf.~Ref.~\refcite{Klebanov:2000me}.
  \item $D3$ branes: A stack of $N$ $D3$ branes has an open-string
        description equivalent to $\mathcal{N}=4$ SYM
        \cite{Witten:1995im}. The stack also forms
        a solution of type IIb supergravity, i.~e.~the low-energy
        limit of type IIb string theory\cite{Duff:1991pea} which
        in the near-horizon limit of the extremal solution reduces to
        the $AdS_5 \times S^5$ metric. This equivalence
        further relates the coupling constant $g$ and number
        of colors $N$ of the gauge theory with the string
        length and coupling $\ell_s$, $g_s$ and the spacetime
        curvature radius $L$ on the gravity side. Specifically,
        \begin{equation}
          g^2 = 4\pi g_s,~~~~~~~~~~g^2 N = \frac{L^4}{\ell_s^4}.
          \label{eq:parameters}
        \end{equation}
  \item The Hawking-Bekenstein entropy associated with the $AdS_5$
        BH of the near-extremal $D3$ brane is equal to the entropy
        of a non-ideal gas in $\mathcal{N}=4$ SYM in the
        strong 't Hooft coupling limit
        \cite{'tHooft:1973jz} $g^2N \rightarrow \infty$;
        cf.~\refcite{Gubser:1996de,Kim:1999sg,Nieto:1999kc} and
        Sec.~2.3 in \refcite{Klebanov:2000me}.
  \item The ratio of shear viscosity $\eta$ to entropy density $s$
        is conjectured to have a lower bound
        given by the value $\eta/s = \hbar/(4\pi)$ which is
        universal to all theories with a gravity dual
        \cite{Kovtun:2003wp}. This bound appears to be
        satisfied by all known liquids.
  \item AdS/CFT based studies (cf.~below)
        predict rapid thermalization on time scales $\sim 1~{\rm fm}/c$
        for quark-gluon plasma produced at Brookhaven's
        Relativistic Heavy Ion Collider (RHIC)
        in agreement with observations
        \cite{Heinz:2004pj}.
\end{itemize}
One of the most valuable properties of the correspondence arises
from Eq.~(\ref{eq:parameters}). The duality is strongly/weakly coupled
and can be applied in both directions:
strong coupling of the gauge
theory $g^2N \gg 1$ corresponds to $L \gg \ell_s$, i.~e.~the
regime of validity of general relativity, while the gauge theory is rather well
understood for small $g^2N$ and may provide information about
the strongly coupled limit of string-theory. We need to bear in mind,
however, that the correspondence applies to cousins of quantum
chromodynamics (QCD) instead of QCD itself. Nevertheless,
the correspondence provides an unprecedented opportunity to
theoretically model
physical systems in the strong-coupling regime of field
theories and, as demonstrated by the above mentioned studies
on shear viscosity and thermalization of quark-gluon plasma,
is capable of making qualitatively and even quantitatively
accurate predictions for physical systems.
Reviews of the AdS/CFT correspondence with varying levels of detail
are given in
\refcite{Aharony:1999ti,Petersen:1999zh,Klebanov:2000me,deBoer:2002da,Horowitz:2006ct,Nastase:2007kj}. For an overview of potential applications
in fluid dynamics, condensed matter physics,
super conductors and super fluidity we refer to
Refs.~\refcite{Son:2007vk,Hartnoll:2009sz,Rangamani:2009xk,Herzog:2009xv,Gubser:2010nc,Hubeny:2010wp}.

Anti-de Sitter space in $d$ dimensions ${\rm AdS}_d$ is the maximally symmetric
solution of the $d$-dimensional vacuum Einstein equations with
cosmological constant $\Lambda<0$
\begin{equation}
  R_{\mu \nu} - \frac{1}{2}Rg_{\mu \nu} + \Lambda g_{\mu \nu}
      = 8\pi G T_{\mu \nu}=0\,.
\end{equation}
It can be represented as the hyperboloid
$X_0^2 + X_{d}^2- \sum_{i=1}^{d-1} X_i^2$ embedded in a $d+1$-dimensional
flat spacetime of signature $--+\ldots+$ with metric
\begin{equation}
  ds^2 = -dX_0^2 - dX_d^2 + \sum_{i=1}^{d-1} dX_i^2\,.
\end{equation}
Transforming to coordinates
\begin{equation}
  X_0 = L \frac{\cos \tau}{\cos \rho},~~~
  X_d = L \frac{\sin \tau}{\cos \rho},~~~
  X_i = L \tan \rho~\Omega_i~~{\rm for}~~i=1\ldots d-1,
\end{equation}
with hyperspherical coordinates\footnote{e.~g.~in $d=5$:
$\Omega_1=\sin \chi \sin \theta \cos \phi$,
$\Omega_2=\sin \chi \sin \theta \sin \phi$,
$\Omega_3=\sin \chi \cos \theta$, $\Omega_4=\cos \chi$ and
$d\Omega_{d-2=3}^2=d\chi^2 + \sin^2 \chi (d\theta^2 + \sin^2\theta d\phi^2)$.}
$\sum _{i=1}^{d-1} \Omega_i^2=1$,
we obtain ${\rm AdS}_d$ in global coordinates with metric
\begin{equation}
  ds^2 = \frac{L^2}{\cos^2 \rho}
      \left( -d\tau^2 + d\rho^2 + \sin^2 \rho d\Omega_{d-2}^2
      \right),
  \label{eq:AdS_globalmetric}
\end{equation}
where $0\le \rho <\pi/2$, $-\pi<\tau \le \pi$. By unwrapping the cylindrical
direction, the range of the time coordinate is often extended to
$\tau \in {\mathbf R}$.

Alternatively, Poincar{\'e} coordinates $(t,z,x^i)$ defined by
\begin{eqnarray}
  && X_0 = \frac{1}{2z}\left[ z^2 + L^2 + \sum_{i=1}^{d-2}(x^i)^2 - t^2
      \right],~~
  X_i = \frac{Lx^i}{z},~~{\rm for}~~i=1\ldots d-2, \nonumber \\
  && X_{d-1} = \frac{1}{2z} \left[z^2 - L^2 + \sum_{i=1}^{d-2}(x^i)^2 = t^2
      \right],~~
  X_d = \frac{Lt}{z},
\end{eqnarray}
give the AdS metric in the form
\begin{equation}
  ds^2 = \frac{L^2}{z^2}
         \left[ -dt^2 + dz^2 + \sum_{i=1}^{d-2}(dx^i)^2 \right]\,,
  \label{eq:AdS_Poincare}
\end{equation}
with $z>0$, $t \in {\mathbf R}$. Poincar{\'e} coordinates cover only
half the hyperboloid (the other half corresponding to $z<0$) and
this patch is often referred to as the {\em Poincar{\'e} wedge}. For its
relation to global AdS see for example
\refcite{Banados:1992gq,Bayona:2005nq,Moschella2005}.

At the AdS boundary $\rho \rightarrow \pi/2$ or $z\rightarrow 0$,
the AdS metric becomes singular. It therefore induces
a conformal class of metrics at the boundary, i.~e.~the induced
metric is determined up to a conformal rescaling only. This
remaining freedom manifests itself in the boundary topology
of the global and Poincar{\'e} metrics. In the limit
$\rho \rightarrow \pi/2$ or $z\rightarrow 0$,
Eqs.~(\ref{eq:AdS_globalmetric}) and (\ref{eq:AdS_Poincare})
become respectively
\begin{equation}
  ds_{\rm gl}^2 \sim -d\tau ^2 + d\Omega_{d-2}^2,~~~~~~
  ds_{\rm P}^2 \sim -dt^2 + \sum_{i=1}^{d-2} d(x^i)^2,
\end{equation}
which can be related by a conformal transformation; cf.~Eq.~(8.17)
in \refcite{Nastase:2007kj}. In consequence, gravity
in global or Poincar{\'e} AdS is related to field theories
on spacetimes of different topology: ${\bf R}\times S_{d-2}$
in the former and ${\bf R}^{d-1}$ in the latter case.

There remains the extraction of physical quantities of the field theory
from the gravitational side. This {\em dictionary} between bulk
and boundary physics is established by the equivalence of the
gravitational action of the bulk taken in the limit of the
boundary and the effective action of the field theory on the
boundary \cite{Maldacena:1997re,Gubser:1998bc,Witten:1998qj}.
This duality involves divergences on both sides, {\em infrared}
divergences on the gravity side due to the singular nature
of the metric on the boundary and {\em ultraviolet} divergences
of the quantum field theory. These anomalies and their
elimination through renormalization have been discussed
in \refcite{Balasubramanian:1999re,deHaro:2000xn,Skenderis:2002wp}.
NR applications have so far focused on the
calculation of the vacuum expectation
values of the field theory's energy momentum tensor
$\langle T_{ij} \rangle$ corresponding to the gravitational sector
in the bulk and we shall only consider this case here.
For a discussion including matter fields see for
example Sec.~5 in \refcite{deHaro:2000xn} as well
as \refcite{Skenderis:2002wp}.

Through the AdS/CFT correspondence, the
vacuum expectation values $\langle T_{ij} \rangle$ of the field theory
are given by the quasi-local Brown-York \cite{Brown:1992br}
stress-energy tensor and thus directly
related to the bulk metric. Following de Haro {\em et al.}
\cite{deHaro:2000xn}, it
is convenient to consider the (asymptotically AdS) bulk metric in
Fefferman-Graham \cite{Fefferman1985} coordinates
\begin{equation}
  ds^2 = g_{\mu \nu} dx^{\mu} dx^{\nu} =
       \frac{L^2}{r^2} \left[ dr^2 + \gamma_{ij} dx^i dx^j \right]\,,
  \label{eq:FeffermanGraham}
\end{equation}
where
\begin{equation}
  \gamma_{ij} = \gamma_{ij}(r,x^i) =
      \gamma_{(0)ij} + r^2 \gamma_{(2)ij} + \ldots
      + r^d \gamma_{(d)ij} h_{(d)ij} r^d \log r^2
      + \mathcal{O}(r^{d+1})\,.
\end{equation}
Here, the $\gamma_{(a)ij}$ and $h_{(d)ij}$ are functions of the
boundary coordinates $x^i$, the logarithmic term only appears for
even $d$ and powers of $r$ are exclusively even up to order $d-1$.
As shown in \refcite{deHaro:2000xn}, the vacuum expectation value
of the CFT momentum tensor for $d=4$ dimensions is then obtained from
\begin{eqnarray}
  \langle T_{ij} \rangle &=&
      \frac{4L^3}{16\pi G} \left\{
      \gamma_{(4)ij} - \frac{1}{8}\gamma_{(0)ij} \left[
      \gamma_{(2)}^2- \gamma_{(0)}^{km} \gamma_{(0)}^{ln}
      \gamma_{(2)kl} \gamma_{(2)mn} \right] \right.
      \nonumber \\
   && \left.
      - \frac{1}{2} \gamma_{(2)i}{}^m\gamma_{(2)jm}
      + \frac{1}{4} \gamma_{(2)ij} \gamma_{(2)}
      \right\} \,,
\end{eqnarray}
and $\gamma_{(2)ij}$ is determined in terms of $\gamma_{(0)ij}$.
The dynamic freedom of the CFT is thus
encapsulated in the fourth-order term $\gamma_{(4)ij}$.
Note that for $r \rightarrow 0$,
the metric (\ref{eq:FeffermanGraham}) asymptotes
to the AdS metric in Poincar{\'e}
coordinates (\ref{eq:AdS_Poincare}). The relation between
the gravitational and CFT energy-momentum tensor for the
case of global AdS is discussed in \refcite{Balasubramanian:1999re};
see also \refcite{Bantilan:2012vu}.

Many applications of the AdS/CFT correspondence are concerned with
the equilibration of matter in heavy-ion collisions at the
RHIC or LHC and in particular its rapid thermalization
\cite{Shuryak:2003xe,Shuryak:2004cy,Muller:2008zzm}.
While the quark-gluon plasma
generated in the collisions is far from equilibrium at
early stages, its behavior appears to be well described by
hydrodynamics after time scales of the order $~1~{\rm fm}/c$.
This process is in principle governed by QCD but results indicate that
many physical aspects can be studied in the framework of
$\mathcal{N}=4$ SYM through the gauge-gravity duality. Small perturbations of
a static system in thermal equilibrium are known to decay
exponentially fast and correspond to quasi-normal modes on
the gravity side \cite{Horowitz:1999jd} and, as we shall discuss,
numerical studies in the context of the AdS/CFT correspondence
yield a similar picture for far-from equilibrium configurations.

The use of NR methods to study these
processes in AdS/CFT has been pioneered by Chesler \& Yaffe
\cite{Chesler:2008hg,Chesler:2009cy} who evolve an
anisotropic source on the AdS boundary switched on after
a short time using a characteristic approach based
on ingoing Eddington-Finkelstein coordinates on the
Poincar{\'e} patch of AdS. Their numerical scheme is
reduced to an effective ``1+1'' scheme by assuming boost invariance
as well as rotational and translation symmetry in the transverse
plane. Their boundary data generate gravitational waves which
propagate into the bulk and lead to formation of a BH.
They extract the energy momentum tensor of the CFT dual
and follow the time evolution of the energy density as well
as the transverse and longitudinal pressure components.
Isotropization of the pressure occurs
on time scales inversely proportional to the local temperature at
the onset of the hydrodynamic regime which translates into
an isotropization time of $~0.5~{\rm fm}/c$ assuming a temperature
of $350~{\rm MeV}$. Using the same setup,
Chesler \& Teaney \cite{Chesler:2011ds} use different definitions
of a ``temperature'' based on the energy density and on two-point
functions. These two versions start agreeing after
a time $~1~{\rm fm}/c$ coincident with the isotropization of
transverse and longitudinal pressure.
Wu \& Romatschke \cite{Wu:2011yd} employ a similar approach
to collide two superposed shockwaves in a boost invariant
approximation and
find that the late-time behavior of the energy density
is given by a hydrodynamic description involving a scale
parameter determined by the initial apparent horizon area.

A comparison of the fully non-linear numerical results with
predictions from the linearized close-limit approximation \cite{Price:1994pm}
was performed by Heller {\em et al.} \cite{Heller:2012km}.
Instead of sourcing the anisotropy through a boundary term,
they prescribe anisotropic data on an initial null hypersurface
extending through the bulk. Their results confirm the short
isotropization times $\sim 1~{\rm fm}/c$ and find the linear approach to
reproduce these values within $~20~\%$ even for large initial
anisotropies.

A numerical scheme based on the ADM formalism of the Einstein
equations was developed by Heller {\em el al.}
\cite{Heller:2011ju,Heller:2012je} who emphasize that thermalization,
when defined as the onset of the hydrodynamical
regime\footnote{see Eq.~(5) in Ref.~\refcite{Heller:2011ju} for their
precise definition}, may differ
from isotropization. By evolving boost-invariant, transversely
homogeneous plasmas with various different initial conditions,
they find that a hydrodynamic
description may well be applicable when the pressure is still
anisotropic.

In Ref.~\refcite{Chesler:2010bi}, Chesler \& Yaffe relax their
symmetry assumptions to translation invariance in the transverse
direction which effectively constitutes a 2+1 numerical
scheme. Their characteristic formulation works well in this
scenario without showing any signs of formation of caustics
and enables them to model heavy-ion collisions by colliding
two shock waves. Specifically, they consider a single shock-wave
solution in Fefferman-Graham coordinates, superpose two of those
and transform the result to ingoing Eddington-Finkelstein coordinates.
For stability purposes, they introduce a small energy offset on
the CFT side which generates an apparent horizon in the gravity dual
above the Poincar{\'e} boundary to absorb steep gradients encountered
in the metric functions deep in the bulk. By comparing the numerically
determined pressure components with hydrodynamical predictions, they
confirm the picture of rapid isotropization obtained in scenarios
of higher symmetry. Translated into values for gold ion collisions
at the RHIC, they observe isotropization about $0.35~{\rm fm}/c$
after their shock waves start overlapping.

A numerical code based on Cauchy type evolutions of
asymptotically global AdS spacetimes using
the GHG scheme
has been developed by Bantilan {\em et al.} \cite{Bantilan:2012vu}.
By assuming an $SO(3)$ symmetry, they reduce their computational
domain to 2+1 dimensions. Initial data is specified in the
form of a localized scalar field which promptly collapses to a
BH with a highly distorted horizon and settles down to
a stationary configuration through quasi-normal ringdown.
Whereas the lowest ringdown modes agree with linearized predictions,
higher angular modes exhibit significant coupling due to
non-linear effects. The dual stress-energy tensor of the CFT is
mapped from the global ${\bf R}\times S^3$ AdS boundary
to a Minkowski background and
found to evolve in agreement with that of a thermalized SYM fluid
from the start of the simulations. Their work furthermore discusses
in detail a number of regularization procedures required to
obtain a numerically stable framework.

A very recent paper by Adams {\em et al.} \cite{Adams:2012pj}
presents a first numerical investigation of superfluid
turbulence in 2+1 dimensions corresponding to
a 3+1 dimensional asymptotically AdS gravity dual.
They observe energy injected at long wavelengths to cascade down
to short length scales, in contrast to the behavior of normal
fluid turbulence in 2+1 dimensions.

\section{Conclusions}
\label{sec:conclusions}

The vast majority of the work reviewed here has been generated
during the last four years; NR in higher dimensions
is a very young research field and likely has only scratched the
surface of a wealth of possible applications. Nevertheless,
a number of impressive results have demonstrated its vast potential.
Discoveries such
as the formation of naked singularities in black strings strengthen
the belief that higher-dimensional gravity differs from its
four-dimensional counterpart in more than a quantitative manner.
The same holds for the asymptotic structure of the spacetimes
under consideration. The instability of asymptotically
anti-de Sitter spacetimes to BH formation under
arbitrarily small perturbations is in stark contrast to the stability
of asymptotically flat or de Sitter
spacetimes\cite{Christodoulou:1993uv,Friedrich:1986}.

One of the remarkable properties of higher-dimensional gravity
is its connection with other areas of physics. The holographic
principle and the AdS/CFT correspondence in particular
have opened a path towards understanding a variety of high-energy
physics phenomena through the study of 4+1 gravity.
This facilitates direct contact with experiments such as heavy-ion
collisions performed at the RHIC. At the same time, parton-parton
collisions at TeV energies at the LHC have provided for the first
time a direct experimental test of the existence of extra dimensions.

In spite of the great success of NR in recent years, a great deal
of work remains to be done. Numerical studies in $d$ dimensions
have as yet been restricted to systems of considerable symmetry.
This is largely a consequence of computational requirements which
increase enormously with each extra dimension. Continuous
improvements of computer technology should, however, make possible
in the foreseeable future the study of generic systems at least in $d=5$,
opening up a much wider class of problems in the context of the
AdS/CFT correspondence and the stability analysis of BHs.

Even within the context of symmetric systems, many challenges are
awaiting the community. Trans-Planckian scattering is rather well-understood
in $d=4$. Connection with experiment, however, requires extension of
these results to higher $d$. Here, progress has been obstructed by
stability issues of the numerical codes, calling for further investigation
of the numerical formulations and gauge conditions.
Many open questions also remain about the stability of BHs in
higher dimensions. Here we note in particular that BHs are no
longer restricted to spherical topology and even stationary multiple-BH
solutions are known. It is unclear how present numerical frameworks
are able to handle black rings or saturns, although the relatively
straightforward extension of NR studies to black strings may be
encouraging. Numerical codes developed for the AdS/CFT correspondence
have as yet been tailored to rather specific problems. It would be
desirable to have a generic framework effectively representing
a numerical AdS/CFT laboratory. Given the enormous range of BH systems
in the context of the correspondence, this is likely going to be
a tremendous challenge, albeit one of large scientific
potential.

In summary, numerical relativity in higher dimensions is a
highly active and rapidly growing field and it will be very exciting
to see its development over the next years.

\section*{Acknowledgments}

The author thanks
E.~Berti,
V.~Cardoso,
R.~Emparan,
P.~Figueras,
L.~Gualtieri,
C.~Herdeiro,
D.~Mateos,
A.~Nerozzi,
H.~Okawa,
H.~S.~Reall,
C.~F.~Sopuerta,
H.~Witek
and M.~Zilh{\~a}o
for many fruitful discussions.
This work was supported by the Ram{\'o}n y Cajal Programme and Grant
FIS2011-304145-C03-03 of the Ministry of Education and Science of Spain.
This work was further supported by
the FP7-PEOPLE-2011-CIG Grant CBHEO-293412,
the FP7-PEOPLE-2011-IRSES Grant NRHEP-295189,
the ERC Starting Grant DyBHo-236667,
the PRACE DECI-7 Grant ``Black hole dynamics in metric theories of gravity'',
the NSF Grant No.~PHY-090003 through XSEDE,
Grants AECT-2012-3-0011 through RES/BSC
and
ICTS-234 through CESGA.
Part of this work was performed with the support of the Cosmos System,
part of the DiRAC HPC Facility funded by STFC and BIS,
and the Baltasar System of IST-CENTRA.

\appendix

\bibliographystyle{ws-ijmpd}


\end{document}